# Optimal Amplitude Multiplexing of a Series of Superconducting Nanowire Single Photon Detectors

**Fabio Chiarello, Roberto Leoni, Francesco Martini, Francesco Mattioli, Alessandro Gaggero**

Istituto di Fotonica e Nanotecnologie, CNR-IFN, Via del Fosso del Cavaliere, 100, 00133 Rome, Italy

Integrated arrays of Superconducting Nanowire Single Photon Detectors (SNSPDs) have shown capabilities such as Photon Number Resolution, single photon imaging and coincidences detection, and can be effectively used also in other different applications related to quantum optics. The growing complexity of such applications requires the use of multiplexing schemes for the simultaneous readout of different detectors. A simple multiplexing scheme can be realized by arranging a series of SNSPDs elements, shunted by appropriate resistances. The goal of this work is to investigate and optimize this scheme, developing a general method able to identify the optimal sets of shunting resistences for any different application. The methodology obtained is very general, and can be extended to other detection systems.

## 1. Introduction

Superconducting Nanowires Single Photon Detectors (SSPD or SNSPD) [1] are fast and sensitive switching detectors particularly suitable for single photon sensing in the visible and near-infrared range [2]. A SNSPD is basically a superconducting nanowire biased by a constant current just below its critical value, arranged in order to maximize the optical coupling with the incident photons (for example, it is patterned in meanders) and shunted by a resistance. In such detectors the absorption of a photon causes the transition of the superconducting wire into the normal state, followed by its self-reset to the superconducting state, ensured by the switching of the current from the meander to the shunting resistance. Because of these processes, the absorption of a single photon produces an output voltage pulse that can be easily detected. Due to this threshold mechanism, the output pulse is independent of both the energy (as long as it is enough to trigger an output pulse) and the number of photons absorbed.

In several applications many SNSPDs have to be simultaneously operated. For example, SNSPDs can be used as pixels in a sensor array for single photon imaging [3,4], as photons number counters [5], as coincidence detectors, and so on. SNSPDs are the only detectors [6] that showed on-chip integration feasibility with outstanding performance in terms of detection efficiency, dark count rate, and timing resolution in the infrared wavelength range [7-13]. Several complex experiments have been performed or proposed that exploit photonic integrated circuits (PICs), including boson sampling, quantum walk, and quantum simulation [14-18]. The growth in the number of detectors requested by the aforementioned applications and experiments causes an increasing complexity that can be addressed only by exploiting various multiplexing strategies. In addition, a large number of RF cable connections inevitably increases the reachable base temperature because of the limited thermal budget of the cryostat. Several multiplexing schemes based on different approaches have been proposed [19-22]. However, SNSPDs allow a simple and effective amplitude multiplexing scheme, obtained by arranging in series different detectors biased by the same constant current and read out by measuring the overall voltage of the entire series [23]. An accurate design of the system enables the identification and discrimination of the states only by a single measurement of the overall voltage, with a great simplification in electronics and design.

This work's main goal is to study and optimize the multiplexing scheme by considering different tasks and conditions, with the identification of requirements and limitations for each particular case.

The results are not limited to the case of SNSPDs but can be easily extended to the optimal multiplexing of many different kinds of systems with analog characteristics, i.e. where a single output is produced by the overall state of many switching detectors.

## 2. Array of SNSPDs

A single SNSPD can be schematized by three lumped elements (FIG. 1 Scheme of a single SNSPD.FIG. 1): an inductance $L$ in series with a switching resistance, given by $\alpha \cdot R_N$ (where $R_N$ is the normal resistance of the wire and $\alpha = 0/1$ is a binary value indicating its non-switching/switching state), arranged in parallel with a shunting resistance $r$ (with $r \ll R_N$), all biased by a constant current $I$.

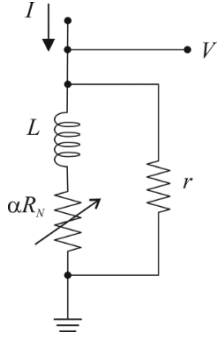

*FIG. 1 Scheme of a single SNSPD.*

For our purposes we restrict the study to the stationary limit, considering only equilibrium conditions reached after fast transients and assuming only real impedances. In this limit the overall resistance is given by the parallel $R = \alpha R_N \cdot r/(r + \alpha R_N)$ (note that for $r \ll R_N$ it is $R \cong \alpha \cdot r$), and the overall output voltage is $V = R \cdot I$.

Let us now consider a series of $n$ different SNSPDs (FIG. 2), each with normal resistances $R_{N,k}$ (the index $k = 1..n$ indicates the $k^{th}$ detector), inductances $L_k$ and shunting resistances $r_k$. The switching state of each single detector $k$ in the series can be expressed by a binary digit $\alpha_k$ ("0" if not switched, "1" if switched), so that the state of an array of $n$ detectors is described by an ordinated sequence of $n$ binary digits, with $2^n$ possible states. The resistance of the $k^{th}$ detector is $R_{Det,k} = r_k \cdot \alpha_k R_{N,k}/(r_k + \alpha_k R_{N,k})$ (it is $R_{Det,k} \cong \alpha_k r_k$ for $r_k \ll R_{N,k}$), and the overall resistance of the entire series is $R = \sum_{k=1}^{n} R_{Det,k}$.

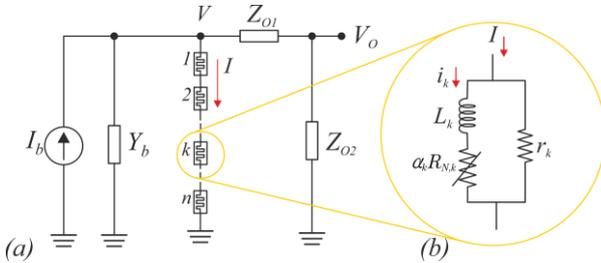

*FIG. 2 Multiplexing scheme with a number n of SNSPDs in series together with a bias and readout network (a). Scheme of the kth SNSPD in the series (b).*

In order to describe the current bias and the voltage readout in a more realistic way, we introduce a general input admittance $Y_b$ and two output impedances $Z_{O1}$ and $Z_{O2}$ (see Fig.2a), describing different effect such as source and amplifier impedances, wires, on-chip lines, and so on. The overall bias/readout admittance is therefore $Y = Y_b + 1/(Z_{O1} + Z_{O2})$. The current and the voltage across the entire SNSPDs' series are respectively:

$$I = I_b - Y \cdot V \qquad (1)$$

$$V = I_b \frac{R}{1 + YR}$$

while the output voltage is $V_O = V \, Z_{O2}/(Z_{O1} + Z_{O2})$.

### 3. Optimal multiplexing

The array of detectors described above can be used for different tasks and applications. For example, it can be used as a counter for Photon Number Resolving (PNR) [5]. In this case it is just required to count the number of detectors switching simultaneously, disregarding any information on which of them is switched. The system can also be used as single-photon pixels-array detector; in this case it is just necessary to identify a single switched detectors at a time. In other cases, the coincident arrival of two or more photons must be detected. Each possible application presents different requirements, and it is possible to design multiplexing schemes that are optimized accordingly. This will be discussed in detail in the following.

As introduced before, for $n$ detectors there are $2^n$ possible switching combinations. Each case can be identified by an index $j$ and described by an array of switching states $\{\alpha_{k,j}\}_{k=1}^{n}$, where $\alpha_{k,j}$ indicates the switching state (0 or 1) of the $k^{th}$ detector for the $j^{th}$ case. The total series resistance $R_j$ is associated to each possible case, corresponding to an output overall voltage $V_{O,j}$. The goal of the multiplexing scheme is to use only the single measurement of the output voltage $V_O$ in order to discriminate between the possible cases of interest. If $\delta V$ is the voltage resolution (the minimal difference of voltage that it is possible to discriminate related, as usual, to the Signal-to-Noise Ratio) the discrimination between two different states $i$ and $j$ is possible only if $|V_{O,i} - V_{O,j}| \geq \delta V$.

In order to clearly illustrate the concept of optimal multiplexing we start by considering the simplest condition, with perfect bias and readout ($Y_i = Z_{O1} = 0$ and $Z_{O2} \to \infty$, so that $Y = 0$, $I = I_b$ and $V = V_O$), and perfect switching ($r_k \ll R_{N,k}$ for any $k$), so that the output voltage for a particular case $j$ is just $V_{O,j} = I_b \sum_k \alpha_{k,j} r_k$. In this simple limit, the identification of any possible case (both number of photons and positions of each of them) can be obtained by performing a kind of "digital-to-analog conversion" of the switching state vector into the output voltage, by choosing a set of shunting resistances with values $r_k = \delta R \cdot 2^{k-1}$, with $\delta R = \delta V / I_b$, so that it is $V_{O,j} = \delta V \sum_k \alpha_{k,j} \cdot 2^{k-1}$. This scheme presents an exponential growth of the involved resistances with the number of detectors, which severely limits this number. Note that it is $V_{O,j+1} - V_{O,j} = \delta V$ (strict equality), and it is simple to verify that any other possible choice of $r_k$ will involve the use of greater resistances, worsening the condition. In this sense, we are considering an optimal multiplexing. In more realistic cases ($Y \neq 0$ and/or $r_k$ comparable to $R_{N,k}$) the situation is worsen.

Fortunately, in general, it is not necessary to discriminate between all the $2^n$ possible cases. This is because, in a specific application, there can be different cases that are equivalent from the point of view of the detection. For example, in the PNR application it is just necessary to detect the number of switched detectors and not which of them are switching. Therefore there will be many different cases giving the same result. We indicate as "equivalence class" a collection of cases that can be considered equivalent from this point of view. In a specific application, it is just required to discriminate the belonging of a particular case to one of these classes. For example, returning to the PNR application, there are only $n+1$ possible equivalence classes: no incident photons (0), one single photon (1), two photons (2), and so on up to $n$ photons detected ($n$). This can relax the exponential growth and the relative limitation but requires detailed considerations on a case-by-case basis.

Let us consider some particular application with $m$ possible equivalence classes (with $m \leq 2^n$), and use a greek index, for example $\beta = 1, .., m$, to indicate these different classes. In an ideal, optimal working condition, we would like to have a unique, well defined and distinguishable value of $V_{O,\beta}$ for each different class $\beta$, corresponding to a different total series resitance $R_\beta$. To discriminate between two of these classes it must be $|V_{O,\beta} - V_{O,\gamma}| \geq \delta V$.

By sorting the values $V_{O,\beta}$ in ascending order and using eq.1 we can rewrite the condition for the discrimination:

$$\frac{R_{\beta+1}}{1+YR_{\beta+1}} - \frac{R_\beta}{1+YR_\beta} \geq \delta R \qquad (2)$$

with resolution $\delta R = \delta V(Z_{O1} + Z_{O2})/(I_b Z_{O2})$. Eq.2 can be rearranged in a more convenient form:

$$R_{\beta+1} \geq R_\beta + \delta R \frac{(1+YR_\beta)^2}{1 - Y\delta R - Y^2 \delta R\, R_\beta} \qquad (3)$$

By considering only the simple equality, and by setting $R_0 = 0$ (the class corresponding to the absence of switchings), eq.3 becomes a recurrence equation with closed solutions:

$$R_\beta = \frac{\beta\, \delta R}{1 - \beta\, Y\, \delta R} \qquad (4)$$

Once defined the classes of equivalent cases for a desired application, and fixed the relations between the resistances of the single detectors and the total resistance of each class, eq.4 allows to determine the optimal set of values.

## 4. Optimal multiplexing for ideal bias and readout

Let us first consider the ideal case with $Y = 0$ and $r_k \ll R_{N,k}$. Eq.2 becomes:

$$R_{\beta+1} - R_\beta = \delta R. \qquad (5)$$

In this case it is simple to study optimal conditions for the different applications.

### 4.1 Photon Number Resolving detection

As previously noticed, for the photon number resolving detection we have $m = n + 1$ different classes, corresponding to the number of detectable photons (from 0 to $n$ photons). This number can be directly used as the class index $\beta$. Since it is not necessary to determine which SNSPDs are switched, they can be considered all identical, with parameters $r_k = r$, $L_k = L$ and $R_{N,k} = R_N$. Therefore, the total resistance for the class with index $\beta$ (remember that $\beta$ is the number of switched detectors) is simply $R_\beta = \beta\, r$. From eq.5, we obtain the (foreseeable) result for the optimal case: the shunting resistances must be equal to $r_k = \delta R$ for any $k$.

### 4.2 Single Photon pixels array detection

In this case, we want to identify which detector is switched because of the incidence of a single photon. Also in this case we have $m = n + 1$ classes (no detections, or one of the $n$ detectors switched), with the index $\beta$ indicating the switched detector (so that in this case the indices $\beta$ and $k$ are equivalent), or 0 in case of no switching. We have therefore $R_0 = 0$ and $R_\beta = r_\beta$ for $\beta > 0$. The optimal solution is given by solving the recurrence equation obtained from eq.5:

$$\begin{cases} r_0 = 0 \\ r_{\beta+1} = r_\beta + \delta R \end{cases} \qquad (6)$$

where it has been introduced a "fictious" resistance $r_0$ corresponding to the zero case. The solution of eq.6 gives the correct set of resistances: $r_k = k\, \delta R$. Note that in this case the growth is only linear and not exponential.

### 4.3 Two Photons detection

In this application it is required the identification of the position of (at maximum) two possible simultaneous switchings. In this case we have 1 class with no switchings, $n$ classes with a single switching, and a number of couples of switchings given by the binomial factor $\binom{n}{2}$, so that the total number of cases is $m = 1 + n + \binom{n}{2}$. Let us dispose the series of shunting resistances $r_k$ sorted in ascending order, and formulate the request that the switching of a single detector with resistance $r_{k+1}$ must be distinguishable by the worst previous case, corresponding to the simultaneous swithcing of the previous two detectors with larger resistances. This gives the following recurrence equations:

$$\begin{cases} r_0 = r_{-1} = 0 \\ r_{k+1} = r_k + r_{k-1} + \delta R \end{cases} \quad (7)$$

were we have, again, introduced fictious resistances $r_0$ and $r_{-1}$. The solution of eq.7 gives the set of optimal shunting resistances: $r_k/\delta R = (3F_k + L_k)/2 - 1$, where $F_k$ and $L_k$ indicate respectively the Fibonacci and Lucas numbers with index $k$. The first six terms are given by the series $r_k/\delta R = \{1, 2, 4, 7, 12, 20, ...\}$. For larger values of $k$ the growth is approximately given by a power law, $r_k \sim 1.62^k$.

*4.4 Detection of three or more Photons*

For the simultaneous detection of three or more photons we can procede in a similar way, obtaining the following recurrence equations:

$$\begin{cases} r_k = 0 \quad (for\ k \leq 0) \\ r_{k+1} = \sum_{l=0}^{n_c-1} r_{k-l} + \delta R \quad (for\ k > 0) \end{cases} \quad (8)$$

were $n_c$ is the number of maximum coincidences to be detected (3 photons, 4 photons and so on), with $n_c \leq n$. Note that for $n_c = n$ there is a closed solution, the previously found $r_k = \delta R\ 2^k$.

## 5. Optimal multiplexing in more general cases

Let us now consider more general cases, in particular for $Y \neq 0$ (realistic bias and readout) and with no assumptions on $r_k$ and $R_{N,k}$ (finite normal resistances). Concerning the realistic bias and readout, we observe that the requirement on the positivity of the denominator in eq.4 fix a limit to the maximum number of classes that can be discriminated, $m < m_L$, with limit $m_L = \beta_{max} = 1/(Y\delta R)$.

Concerning the normal resistances, we consider that the resistance of the $k^{th}$ SNSPD can switch between 0 and the parallel $r_{P,k} = r_k \cdot R_{N,k}/(r_k + R_{N,k})$. Starting from these general considerations, let us examine again the different applications.

*5.1 Photon Number Resolving detector*

Once again, in the PNR application all the SSPDs can be considered indistinguishable, with equal resistances $r_k = r$, and with parallel of the shunting resistance with the normal one given by $r_P = r \cdot R_N/(r + R_N)$. In this case the goal is to count the number of switched detectors. By assuming a number of detectors $n < m_L = 1/Y\delta R$, and considering that the total resistance corresponding to a number $\beta$ of switched detectors is $R_\beta = \beta\ r_p$, one obtains from eq.4:

$$\frac{r_P}{\delta R} = \frac{1}{1 - n/m_L} \quad (9)$$

and, by inverting the relation between $r$ and $r_P$, $r = r_P R_N/(R_N - r_p)$, it is:

$$\frac{r}{\delta R} = \frac{1}{1 - n/m_L - \delta R/R_N} \quad (10)$$

Note that for $Y \to 0$ and $R_N \to \infty$ this result is equivalent to that obtained in paragraph 4.1.

As an example, we consider the case reported in ref.[5]. In this case we have $Y = 1/(50\Omega)$ and an extimated $\delta R \simeq 2\ \Omega$, so that $m_L \simeq 25$. The used number of $n = 24$ detectors is therefore optimum and, for $R_N = 1.6\ k\Omega$, it must be $r \simeq 60\ \Omega$, close to the used value of 70 $\Omega$

*5.2 Single Photon pixels array detection*

In this case (see ref.[23]) there is only a single detector switching at a time (or no detectors switching) so that the total resistance is equal to the resistance of the switched element: $R_\beta = r_{P,\beta}$. It is possible to use directly eq.4 and the relation $r_k = r_{P,k} R_N/(R_N - r_{p,k})$ to obtain:

$$\frac{r_k}{\delta R} = \frac{k}{1 - k/m_L - k\ \delta R/R_N} \quad (11)$$

Note that for $Y \to 0$ and $R_N \to \infty$ this result is equivalent to that obtained in paragraph 4.2.

In tab.1 we consider a realistic example with $\delta R = 2\ \Omega$ in three different cases: (A) ideal case with $Y = 0\ S$ and $R_{N,\beta} \to \infty$; (B) imperfect bias and readout, with $Y = 1/(50\Omega)$, but $R_{N,\beta} \to \infty$; (C) imperfect bias and readout and finite normal resistances, with $R_{N,\beta} = 1\ k\Omega$ and $Y = 1/(50\Omega)$. Note that in cases (B) and (C) it is $m_L = 25$, so that it must be $k < 25$. Moreover, in case (C) there is a lower limit ($k < 24$) due to the last term in the denominator of eq.11.

| $k$ | $r_k^A(\Omega)$ | $r_k^B(\Omega)$ | $r_k^C(\Omega)$ |
|---|---|---|---|
| **0** | 0.00 | 0.00 | 0.00 |
| **1** | 2.00 | 2.08 | 2.09 |
| **2** | 4.00 | 4.35 | 4.37 |
| **3** | 6.00 | 6.82 | 6.86 |
| **4** | 8.00 | 9.52 | 9.62 |
| **5** | 10.00 | 12.50 | 12.66 |
| **6** | 12.00 | 15.79 | 16.04 |
| ... | ... | ... | ... |
| **22** | 44.00 | 366.67 | 578.95 |
| **23** | 46.00 | 575.00 | 1352.94 |
| **24** | 48.00 | 1200.00 | - |

*Table 1. Shunting resistances optimized for single photon pixels array detection, with $\delta R = 2\ \Omega$ and three different conditions: (A) $Y = 0\ S$, $R_{N,\beta} \to \infty$; (B) $Y = 1/(50\Omega)$, $R_{N,\beta} \to \infty$; (C) $Y = 1/(50\Omega)$, $R_{N,\beta} = 1\ k\Omega$.*

## 5.3 Detection of two or more Photons

In this case it is possible to proceed as in paragraph 4.3 and 4.4, considering the possibility to have simultaneously up to $n_c$ detectors switched. We dispose in ascending order the shunting resistances $r_\beta$ (and, consequently, the parallel resistances $r_{P,\beta}$) and use eq. 3 with appropriate assignments. The total resistance corresponding to the switching of the single element $\beta + 1$ is $R_{\beta+1} = r_{p,\beta+1}$, while the total resistance for the previous worst case, the simultaneous switching of the previous $n_c$ elements, is $R_\beta = \sum_{l=0}^{n_c-1} r_{P,\beta-l}$. Their difference must be greater than $\delta R$, so that:

$$r_{P,\beta+1} = R_\beta + \delta R \frac{(1+YR_\beta)^2}{1 - Y\delta R - Y^2 \delta R\, R_\beta} \quad (12)$$

By solving the recurrence eq. 12 one obtains the set of shunting resistances from $r_\beta = r_{P,\beta} R_N/(R_N - r_{p,\beta})$.

In table 2 it is considered an example with optimal resistances for the detection of two photons for the three cases presented in Tab.1. Once again, in cases (B) and (C) it is $m_L = 25$, so that it should be $k < 25$, but we observe a more severe limit to the number of possible cases ($k \leq 9$ in case (B) and $k \leq 8$ in case (C)).

| $k$ | $r_k^A(\Omega)$ | $r_k^B(\Omega)$ | $r_k^C(\Omega)$ |
|---|---|---|---|
| 0 | 2.00 | 2.08 | 2.09 |
| 1 | 4.00 | 4.35 | 4.37 |
| 2 | 8.00 | 9.10 | 9.18 |
| 3 | 14.00 | 16.84 | 17.13 |
| 4 | 24.00 | 30.85 | 31.83 |
| 5 | 40.00 | 55.97 | 59.29 |
| 6 | 66.00 | 103.64 | 115.62 |
| 7 | 108.00 | 201.84 | 252.89 |
| 8 | 176.00 | 446.75 | 807.51 |
| 9 | 286.00 | 1533.68 | - |
| 10 | 464.00 | - | - |

*Table 2. Shunting resistances optimized for two photons pixels array detection, with $\delta R = 2\, \Omega$ and the three different conditions: (A) $Y = 0\, S$, $R_{N,\beta} \to \infty$; (B) $Y = 1/(50\Omega)$, $R_{N,\beta} \to \infty$; (C) $Y = 1/(50\Omega)$, $R_{N,\beta} = 1\, k\Omega$.*

## 6. Summary

In this paragraph we summarize the obtained results. The problem consist in finding the optimal set of shunting resistence $r_k$, with values minimal but still sufficient to distinguish different equivalent classes. We consider different possible applications:

- Photon Number Resolving: count the number of switched detectors, with no information on which detectors are switched.
- Single photon pixel array: identification of which detector has switched (single detector).
- Detection of two or more photons: identify the switched detectors, up to a given number of coincidences.
- Full detection: detection and identification of any possible condition.

The value $\delta R$ is the "resistance resolution". The quantity $m_L = (Y\delta R)^{-1}$ gives a limit to the number of detectors, with $Y$ being the total admittance of the bias /readout circuitry. $R_N$ is the normal resistance of a single detector (supposing identical detectors). In the ideal condition it is $Y = 0\, S$ and $R_N \to \infty$. In tab.3 there are reported the results.

| Application | Ideal condition | Non ideal condition |
|---|---|---|
| Photon Number Resolving | $r_k = \delta R$ | $r_k = \dfrac{\delta R}{1 - n/m_L - \delta R/R_N}$ |
| Single Photon Detection | $r_k = k\, \delta R$ | $r_k = \dfrac{k\, \delta R}{1 - k/m_L - k\, \delta R/R_N}$ |
| Detection of more photons | See eq. 8 | See eq. 12 |
| Full detection | $r_k = 2^k \delta R$ | |

*Table 3. Summary of the optimal set of shunting resistances for different applications, under ideal or non ideal conditions.*

In figure 3 there are plotted the growths of resistances $r_k$ for the different examples considered in this work: dotted lines for the ideal case (with $\delta R = 2\, \Omega$) and continuous lines for the non-ideal ones with $Y = 1/(50\Omega)$ and $R_N = 1\, k\Omega$; blue lines for the Single Photon Pixel Array detection, red lines for the two photons coincidence, green line for the full detection (only in the ideal case).

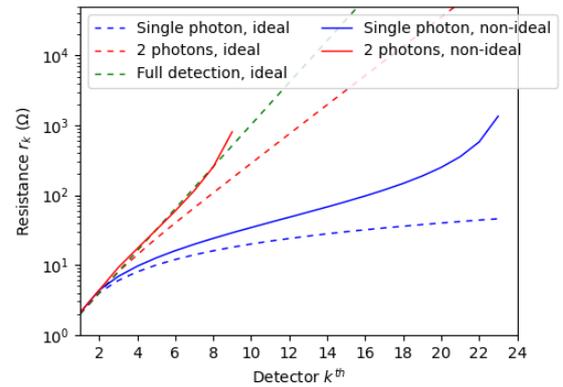

*FIG. 3. Growths of the shunting resistances $r_k$ for the different examples presented in this work: dashed lines for ideal cases, continuous lines for non-ideal cases with $Y = 1/(50\, \Omega)$ and $R_N = 1\, k\Omega$.*

## 7. Conclusions

We have considered the amplitude multiplexing of an array of SNSPDs in series, and developed a general methodology for the definition of the optimal set of shunting resistances, depending on the considered application. The presented method is very general, and can be be extended to different typologies of detectors.

## Acknowledgements

This work is supported by the Next GenerationEU projects Rome Technopole, National Quantum Science and Technology Institute (NQSTI) and by the Italian National Centre for HPC, Big Data and Quantum Computing (HPC).